\title{Incompatibility between 't Hooft's and Wolfram's models of quantum mechanics}
\author{
        Jos\'e Manuel Rodr\'iguez Caballero \\
                External affiliate of the Wolfram Physics Project \\ Director of Caballero Software Inc. \\ Ontario, Canada
}
\date{\today}

\documentclass[12pt]{article}

\usepackage[english]{babel}

\usepackage{blindtext}

\usepackage{amsmath}
\usepackage{hyperref}
\usepackage{lipsum}
\usepackage{amsfonts}
\usepackage{enumitem}
\usepackage{epigraph}
\usepackage{graphicx}
\usepackage{amsthm}

\theoremstyle{definition}

\begin{document}
\maketitle

\begin{abstract}
Stephen Wolfram and Gerard 't Hooft developed classical models of quantum mechanics. We show that the descriptive complexity grows differently as a function of time in each model. Therefore, they cannot describe the same physical system. In addition, we propose an interpretation of the Wolfram model, which shares some characteristics with 't Hooft's model, but which involves a non-computable function.
\end{abstract}

\section{Introduction}
\epigraph{Randomness is the true foundation of mathematics.}{Gregory Chaitin}

The development of classical models of quantum mechanics is limited by what is called no-go theorems. These theorems, initially proved by John von Neumann \cite{von2018mathematical}, prevent the creation of quantum mechanical models satisfying certain assumptions of classical physics. Nevertheless, this does not contain some authors from reinterpreting the experimental proofs of quantum effects and from developing new models beyond the framework of these theorems. Two examples are Gerard 't Hooft's cellular automata interpretation  \cite{t2016cellular}, and Stephen Wolfram's physics project \cite{wolfram2020class}.

Gerard 't Hooft's point of departure is to deny the existence of quantum superpositions. The time is discreet; the smallest unit is an iteration of the universe as a computer system. In his model, there is a special basis of the Hilbert space of the universe, called the ontological basis. The vectors of this base, called ontic states, are intended to represent the states of the universe in reality. Ontic states are postulated to evolve into ontic states after an iteration of the universe. This evolution is deterministic, and it can be simulated in a (classical) computer provided with sufficient memory and time. Because there are many similarities between `t Hooft's model and the model studied by Stephen Wolfram in A New Kind of Science \cite{wolfram2002new}, this article can be interpreted as a comparison between the first and second approaches of Wolfram's fundamental physics, expressed in `t Hooft's theoretical framework (we use the` t Hooft language of ontic states rather than Wolfram's language of hypergraphs, which are the ontic states in his model).

In quantum mechanics\footnote{The Wolfram model is also developed in the direction of the theories of special and general relativity, but we will omit these subjects in the present paper.}, Stephen Wolfram's starting point can be seen as a non-deterministic generalization of `t Hooft's model, known as the multiway system. In this framework, quantum effects are intended to be explained in a way similar to Hugh Everett's many-worlds interpretation \cite{everett2015theory}. Of course, for the case of a single-world, both Wolfram's and Hooft's models are essentially the same. The word ``incompatibility'' in the title is used when the Wolfram model assumes more than one world in the sense described below.

This article aims to show that, assuming more than one world and an aperiodic evolution, the descriptive complexity of the universe perceived by a typical version of an observer in Wolfram's model increases as a linear function of time (generically). In contrast, the same magnitude in `t Hooft's model grows as a logarithmic function of time (generically). By generically, we meant that all the exceptions are negligible in a sense that will be specified in the elaboration of the results. Therefore, the two models cannot describe the same physical system. Nevertheless, we will provide a version of the Wolfram model, called its client-server interpretation, which shares some characteristics of the `t Hooft model, for example, there is only one world, but the experience of the observer is similar to that of the typical version of the observer in Wolfram's model.

\section{Typical observer's experience}

A model of quantum mechanics is \emph{Wolfram-like} if it postulates that an ontic state always evolves into a multi-set of ontic states. A Wolfram-like model is  \emph{$k$-regular} if this multi-set always contains $k$ different elements\footnote{The repetition of elements in the multiset may be related to the probability amplitude of a superposition of ontic states in the model.}. A \emph{`t Hooft-like} model is a $1$-regular Wolfram-like model.

An \emph{observer's experience} in a Wolfram-like model is a sequence of ontic states $|x_1\rangle, |x_2\rangle, |x_3\rangle,...$ such that, for every $n$, the ontic states $|x_n\rangle$ evolves to a multi-set containing $|x_{n+1}\rangle$. An observer's experience in a $2^r$-regular\footnote{For simplicity, we will focus on the case when $k$, in the definition of $k$-regular, is a power of $2$.} Wolfram model is \emph{typical} if, for all $n$ large enough,  the descriptive complexity of $x_n$ (the label of the ontic state) is equal to a constant plus $r n$, and the set of exception is negligible, i.e., the number of elements of the set of integers smaller than or equal to $n$ not satisfying this property, divided by $n$ converges to zero as $n$ goes to infinity. A $2^r$-regular Wolfram-like model is \emph{non-degenerate} if it contains at least one typical observer's experience.

\textbf{Convention.} From now on, we will assume that all $2^r$-regular Wolfram-like models, with $r \geq 1$, mentioned in this paper are non-degenerate\footnote{An example of a $2$-regular Wolfram-like model that is not non-degenerate is $1 \mapsto {2, 3}$, $2 \mapsto {1, 3}$, $3 \mapsto {1, 2}$}. The goal of this convention is to avoid repeating the word ``non-degenerate'' too many times, making the text easier to read.

Notice that in a `t Hooft-like model, there is a typical observer's experience if and only if the system is periodic. For the aperiodic case, for all $n$ large enough, the descriptive complexity of $x_n$ is equal to a constant plus the number of binary digits of $n$.

\section{Logarithmic time}
In the original approach of S. Wolfram and `t Hooft, the time is proportional to the number of system iterations. Under this assumption, an observer Alice, who is integrated into a  $ 2 ^ r $ -regular Wolfram model, can experimentally determine the value of $ r $, provided enough data on the descriptive complexity of her universe as a function of time. For this, Alice must assume that her experience is typical as a consequence of the Copernican principle.

Turbulence expert W. K. George explored the cosmological consequences of reformulating the laws of physics using the logarithm of time instead of time as usual. Inspired by this idea, we will consider the possibility that an observer of a model of type `t Hooft can perceive the time as proportional to the logarithm of the number of system iterations. An example of such an observer is the red-green color-blind daemon in the following thought experiment.

In a solitary cage, isolated from the rest of the universe, a daemon is imprisoned. The daemon has a stopwatch, expressing time as binary digits, which are square boxes. The red and green boxes represent $0$ and $1$, respectively. The only problem is that the daemon is red-green colour-blind. Therefore, the only information the daemon can get from the stopwatch is the number of digits in the hour. So, as perceived by the stopwatch daemon, time increases logarithmically concerning the number of stopwatch iterations.

\section{Client-server transformation}

The relationship between client and server described below is inspired by the relationship between slow and fast variables, respectively, as used by `t Hooft.

We propose a method for transforming a classical many-worlds model of quantum mechanics (under the Wolfram model) into a single world model, which we call its client-server interpretation. This transformation preserves the experience of a typical observer. The price to pay is the introduction of a non-computable function.

Consider a $2^r$-regular Wolfram-like model for $r \geq 1$. Imagine a server generating the ontic states of the system, one by one, as levels of a $2^r$-ary tree. Each time the server generates an ontic state, it sends it to the client. The client accepts the $n$-th state sent by the server if and only if $n$ is the smallest number of its length in base $2^r$ that cannot be compressed (we assume a fixed Turing machine). Finally, the client uses the sequence of ontic states accepted from the server as the history of the universe. Notice that the time in the universe generated by the client is proportional logarithm of the number of iteration of the server.

Using the Berry paradox, it is easy to prove that the property ``$n$ is the smallest number of its length in base $2^r$ that cannot be compressed'' is non-decidable. We will call the sequence of numbers satisfying this property, the \emph{Chaitin geometric progression} in base $2^r$ (we will omit the base when it is clear from the context).

We have seen that the experience of a typical version of an observer in a $2^r$-regular Wolfram-like model, with $r \geq 1$, is equivalent to the experience of the single version of an observer in its client-server interpretation. Nevertheless, the simplicity of one description can be superior to the other description. For example, consider the case of the $2$-regular Wolfram model where the ontological states are labeled by positive integers and the evolution rules are:
$$
U_0 |x\rangle = |2x\rangle, \quad U_1 |x\rangle = |2x+1\rangle. 
$$

It is easy to show that the evolution rule (in the server) of the corresponding client-server interpretation is just\footnote{The letters c and s stand for client-server.}
$$
U_{\textsf{cs}} |x\rangle = |x+1\rangle. 
$$

\section*{Conclusions}

Unlike the Copenhagen doctrine, where true randomness is imposed in the universe as a postulate, the client-server interpretation gets the same effect from what Gregory Chaitin calls the ``randomness of mathematics'', i.e., randomness intrinsic to the system, not assumed as an independent postulate. The requirement of an oracle for the Chaitin geometric progression is avoided in the $2^r$-regular Wolfram-like model, with $r \geq 1$, in a way similar as in the many-worlds interpretation of quantum mechanics, by assuming multiples versions of the observer and applying the Copernican principle: the typical version experiences randomness, because, in this case, most configurations are incompressible in Chaitin's sense.

As a reaction to the theory of consciousness developed by the school of strong artificial intelligence, Roger Penrose \cite{penrose1989} proposed the existence of non-computational processes in quantum mechanics. This claim seems to contradict the so-called physical Church-Turing thesis, due to Stephen Wolfram \cite{wolfram1985undecidability} and David Deutsch \cite{deutsch1985quantum}. Nevertheless, this contradiction is apparent since the experience of a typical version of an observer in a multiway system is equivalent to the experience of the unique version of an observer in its client-server interpretation (involving non-computable processes). Hence, even if some non-computational process is detected in quantum mechanics as the experience of an observer, it may be the result of the Copernican principle applied to the multiway system.

\section*{Acknowledgments}
The author would like to thank Stephen Wolfram, William K. George and Gregory Chaitin for the interesting exchange of ideas. Also, the author is very grateful to all the members of the Wolfram Physics Project for having contributed every day to the development of this new approach to fundamental physics. Special thanks to Christian Pratt, who collaborated on the initial version of this project.

\nocite{*}

\bibliographystyle{alpha}
\bibliography{mybibfile}

\end{document}